\documentclass[aps,prd,notitlepage,showpacs,nofootinbib,superscriptaddress]{revtex4-1}

\usepackage{graphicx}
\usepackage[utf8]{inputenc} 

\usepackage{amsmath}
\usepackage{amssymb}
\usepackage{float}
\usepackage{comment}

\usepackage{caption}
\usepackage{subcaption}
\usepackage{amsmath}
\usepackage{amssymb}

\usepackage{feynmp-auto}
\usepackage[usenames,dvipsnames]{color}
\usepackage[colorinlistoftodos]{todonotes}
\usepackage[colorlinks=true,citecolor=darkred,urlcolor=darkred, pdfborder={0 0 0}]{hyperref}
\usepackage[normalem]{ulem}

\definecolor{darkred}{rgb}{0.6,0,0}
\usepackage[colorinlistoftodos]{todonotes}

\definecolor{linkcolor}{rgb}{0,0,0.5}
 
\newcommand {\ignore}[1]{}

\def \znbb {$\rm 0\nu\beta\beta$}

\def\gsim{\raise0.3ex\hbox{$\;>$\kern-0.75em\raise-1.1ex\hbox{$\sim\;$}}}
\def\lsim{\raise0.3ex\hbox{$\;<$\kern-0.75em\raise-1.1ex\hbox{$\sim\;$}}}
\def\lfv{lepton flavour violation }

\def\SM{$\mathrm{SU(3)_c \otimes SU(2)_L \otimes U(1)_Y}$ }
\newcommand{\sm}{{Standard Model }}

\usepackage{soul}

\usepackage{feynmp-auto}

\definecolor{mightnightblue}{RGB}{25,25,112}

\definecolor{brown}{rgb}{0.59, 0.29, 0.0}

\newcommand {\black} {\color{black}}

\def\lfv{lepton flavour violation }

\def\SM{$\mathrm{SU(3)_c \otimes SU(2)_L \otimes U(1)_Y}$ }
\def\21{$\mathrm{SU(2)_L \otimes U(1)_Y}$}
\def\lfv{lepton flavour violation }

\def\sm{standard model }

\newcommand{\AddrAHEP}{AHEP Group, Institut de F\'{i}sica Corpuscular 
  (CSIC-Universitat de Val\`{e}ncia), Parc Cientific de Paterna.\\
  C/Catedratico Jos\'e Beltr\'an, 2 E-46980 Paterna (Val\`{e}ncia) - SPAIN}

\begin{document}

\title{\color{BrickRed}  Simplest Scoto-Seesaw Mechanism }

\author{Nicol\'{a}s Rojas}\email{nicolas.rojasro@usm.cl}
\affiliation{Universidad T\'ecnica Federico Santa Mar\'{i}a - Departamento de F\'{i}sica\\
      Casilla 110-V, Avda. Espa\~na 1680, Valpara\'{i}so, Chile}
      
\author{Rahul Srivastava}\email{rahulsri@ific.uv.es}
\affiliation{~\AddrAHEP}

\author{Jos\'{e} W. F. Valle} \email{valle@ific.uv.es, URL: 
  http://astroparticles.es} 
\affiliation{~\AddrAHEP}

\begin{abstract}

By combining the simplest (3,1) version of the seesaw mechanism containing a single heavy ``right-handed'' neutrino with the minimal scotogenic approach to dark matter, we propose a theory for neutrino oscillations. The ``atmospheric'' mass scale arises at tree level from the seesaw, while the ``solar'' oscillation scale emerges radiatively, through a loop involving the ``dark sector'' exchange. Such simple setup gives a clear interpretation of the neutrino oscillation lengths, has a viable WIMP dark matter candidate, and implies a lower bound on the neutrinoless double beta decay rate.

  \end{abstract}


\maketitle

\black
\section{Introduction}

So far the only neutrino mass parameters which have been experimentally measured are the two mass square splittings associated with the ``atmospheric''  and ``solar'' neutrino oscillations~\cite{McDonald:2016ixn,Kajita:2016cak}. Precision measurements at reactor and accelerator-based experiments~\cite{An:2012eh,Abe:2017uxa} further strengthen the three-neutrino oscillation paradigm. The corresponding ratio of squared solar-to-atmospheric mass splittings is found to be~\cite{deSalas:2017kay}
\begin{equation}
  \label{eq:sol-atm}
 \frac{\Delta m_{\rm SOL}^2}{\Delta m_{\rm ATM}^2} = 0.0302_{-0.0010}^{+0.0012}
\end{equation}
at 1~$\sigma$ C.L. This ratio is necessary in order to describe the neutrino oscillation data successfully. The existence of these two different mass scales could be indications that maybe the two mass scales arise from two very different mechanisms. 

Likewise, despite our inability to pin down its nature, we have firm evidence for the existence of cosmological dark matter~\cite{Bertone2005279}. 
Altogether, the lack of neutrino masses and of a viable dark matter candidate constitute two of the main drawbacks of the standard \SM model of particle physics whose minimal gauge structure we tacitly assume here.

Accommodating neutrino oscillation parameters within a minimal tree level type-I seesaw extension of the standard model requires two isosinglet fermion messengers~\footnote{Since singlet fermions, such as ``right-handed'' neutrinos, carry no anomaly, their multiplicity is not fixed~\cite{Schechter:1980gr}. Note that in two-component form all fermions are ``left-handed''. Our $N$ corresponds to what people normally call ``right-handed'' neutrino in 4-component form.} i.e. the (3,2) seesaw, instead of the ``complete'' sequential (3,3) seesaw scenario. Indeed, this (3,2) seesaw possibility, first suggested in~\cite{Schechter:1980gr}, has been recently studied in Ref.~\cite{King:2015sfk}. It is clear however that, in the presence of adequate radiative corrections, the minimal (3,1) type-I seesaw picture could be sufficient. Unfortunately, in such minimal (3,1) type-I seesaw setup the required mass splittings needed to describe solar neutrino oscillations do not emerge automatically. Moreover, the problem of having a suitable dark matter candidate remains unresolved. Here we show that both of these requirements can be met by consistently implementing the scotogenic mechanism proposed by Ernest Ma~\cite{Ma:2006km}.

In this letter we propose a scotogenic extension of the simplest (3,1) seesaw mechanism that provides a very clear interpretation of the two observed oscillation scales and the associated messengers. To first approximation we generate only the atmospheric scale, from the tree level exchange of the single isosinglet neutrino messenger present in the minimal (3,1) type-I seesaw mechanism~\cite{Schechter:1980gr}. At this stage two neutrinos are left massless, so there are no solar neutrino oscillations. We then show that this mass degeneracy is lifted by calculable radiative corrections of the scotogenic type. These involve the one-loop exchange of dark messengers, responsible for generating the solar scale, thus naturally accounting for the observed smallness of solar-to-atmospheric ratio in Eq.~(\ref{eq:sol-atm}). The lightest radiative neutrino mass messenger associated to the solar mass scale provides the observed cosmological dark matter in the form of a thermal weakly interacting massive particle (WIMP).

\section{The minimal scotogenic seesaw }
\subsection{The Yukawa Sector}

Here we describe our minimal scotogenic seesaw mechanism. Our basic template is the ``incomplete'' (3,1) type-I seesaw mechanism proposed in Ref.~\cite{Schechter:1980gr}, containing a single heavy isosinglet neutral lepton, we call $N$. To this we add a dark sector, odd under the assumed ``dark $\mathbb{Z}_2$ parity'', consisting of one scalar $\eta$ and one fermion $f$, indicated in the last two columns of table \ref{tab:Fields}. As mentioned, this theory is consistent, since gauge singlets carry no anomaly. The full Yukawa sector is given by:
\begin{eqnarray}
\mathcal{L_{}} &=& \mathcal{L_{\rm SM}} + \mathcal{L_{\rm ATM}}  + \mathcal{L_{\rm DM,SOL}} \label{eq:tot}
\end{eqnarray}
where $\mathcal{L_{\rm SM}}$ is the \sm Lagrangian and
\begin{eqnarray}
\mathcal{L_{\rm ATM}} &=& -Y_{(N)}^k \overline{L}^k i\sigma_2 H^* N\,  +\, \frac{1}{2} M_N \overline{N^c}N\,  +~h.c. \label{eq:atm}
\end{eqnarray}
induces an effective non-zero tree level neutrino mass, once the electroweak symmetry is broken by the vacuum expectation value of the \sm Higgs. This scale is identified as the atmospheric scale. 
\begin{table}[!h]
\setlength\tabcolsep{0.25cm}
\centering
\begin{tabular}{| c | c | c | c |}
\hline
                      &  $N$   & $\eta$   & $f$ \\
\hline                                    
$\mathrm{SU(2)_L}$    &   1    &    2    &    1     \\
$\mathrm{U(1)_Y}$     &   0    &   1/2   &    0     \\
$\mathbb{Z}_2$        &  $+$   &   $-$   &  $-$     \\
Multiplicity          &   1    &    1    &   1      \\
\hline
\end{tabular}
\caption{\label{tab:Fields} Messengers in the scotogenic extension of the minimal type-I seesaw model, called (3,1) in~\cite{Schechter:1980gr}. }
\end{table}

Note that, by itself, Eq.~(\ref{eq:atm}) leads neither to a dark matter candidate, nor to solar neutrino oscillations. The minimal way to generate the solar neutrino mass scale is to add the two ``dark'' messenger fields, one boson and one fermion, indicated in the last two columns of table \ref{tab:Fields}. Note that the fermion $f$ is intrinsically distinct from the isosinglet $N$ as it is odd under the $\mathbb{Z}_2$ symmetry. 
The Lagrangian responsible for the solar and dark sector is given by
\begin{eqnarray}
  \mathcal{L_{\rm DM,SOL}} &=& Y_{(f)}^k \overline{L}^k i\sigma_2 \eta^* f\, + \frac{1}{2} M_f \overline{f^c}f +~h.c. \label{eq:solar}
\end{eqnarray}
The quantum corrections coming from the ``scotogenic'' loop~\cite{Ma:2006km} will now induce the smaller solar scale.

Note that the last row in Table \ref{tab:Fields} indicates the minimum required field multiplicity ensuring the presence of one atmospheric seesaw messenger and the dark matter candidate, interpreted as the lighter of the messengers responsible for generating the solar neutrino scale. 

In short, our proposal is the simplest scotogenic completion of the (3,1) seesaw model, defined in Eqs.~(\ref{eq:tot},\ref{eq:atm},\ref{eq:solar}). This minimal setup provides a very simple picture where the atmospheric scale arises at the tree level while the solar scale follows from calculable loop corrections, hence matching the observed smallness of one with respect to the other, given in Eq.~(\ref{eq:sol-atm}). These features emerge~\cite{Hirsch:2000ef,diaz:2003as} in more complicated supergravity schemes based upon spontaneous~\cite{Masiero:1990uj} or bilinear breaking of R parity~\cite{Diaz:1997xc}.
However, within the supersymmetric picture, due to the breaking of R-parity, there is no stable WIMP dark matter candidate. In contrast, in the present much simpler setup, by construction the existence of WIMP dark matter is ensured.  

\subsection{The Scalar Sector}

The scalar potential must include a second doublet $\eta$, with the same quantum numbers as the standard model Higgs doublet, but with $\mathbb{Z}_2$-odd parity. It can be written as
\begin{eqnarray}
V & = & -m_H^2 H^\dagger H + m_\eta^2 \eta^\dagger\eta 
+ \frac{1}{2} \lambda_H (H^\dagger H)^2
+ \frac{1}{2} \lambda_\eta (\eta^\dagger\eta)^2
+ \lambda_{3} (H^\dagger H) (\eta^\dagger\eta)
+ \lambda_{4} (H^\dagger\eta) (\eta^\dagger H)
+ \frac{1}{2} \lambda_{5} \Big[(H^\dagger\eta)^2 + h.c.
\Big]
\end{eqnarray}
Since the $\mathbb{Z}_2$ symmetry is exactly conserved, the mixings between the Higgs doublet and $\eta$ are forbidden. The mass spectrum for the components
of the $\eta$ doublet are given by
\begin{eqnarray}
m_{\eta^R}^2 &=& m_\eta^2 + 
\frac{1}{2} (\lambda_{3}+\lambda_{4}+\lambda_{5}) v_\phi^2 \label{eq:etaR} \\
m_{\eta^I}^2 &=& m_\eta^2 + 
\frac{1}{2} (\lambda_{3}+\lambda_{4}-\lambda_{5}) v_\phi^2 \label{eq:etaI} \\
m_{\eta^+}^2 &=& m_\eta^2 + \frac{1}{2} \lambda_{3} \, v_\phi^2~.
\end{eqnarray}
Note that the doublet $\eta$ in the scalar potential is analogous to the one present in the popular inert Higgs doublet model~\cite{LopezHonorez:2006gr}. However, in our present theory we have a new feature associated with the key role of the coupling $\lambda_5$, related to the breaking of lepton number responsible for solar neutrino mass generation~\cite{Merle:2016scw}, a feature absent in the inert Higgs doublet scheme~\footnote{Note that in the present theory neutrinos are massive Majorana fermions, hence lepton number is violated by the Majorana mass term $NN$ responsible for atmospheric neutrino mass generation, as well as by the coupling $\lambda_5$ in the scalar potential, responsible for solar neutrino mass generation.}. 

\section{Neutrino Masses}
\subsection{Tree Level Structure}
\label{sec:tree}

The tree level part of our model gives rise to a neutrino mass matrix $\mathcal{M}_\nu^{ij}$ given by 
\begin{eqnarray}
\mathcal{M}_\nu^{ij} & = &
\left( \begin{array}{cccc}
 0           & 0            & 0            & y^1_{(N)} v \\
 0           & 0            & 0            & y^2_{(N)} v \\
 0           & 0            & 0            & y^3_{(N)} v \\
 y^1_{(N)} v & y^2_{(N)} v  & y^3_{(N)} v  & M_N \\
 \end{array} \right)
\label{tree-part}
\end{eqnarray}
in the basis $(L^k, N)^T$. This clearly has a projective structure:
\begin{eqnarray}
\mathcal{M}_{\nu~\rm TREE}^{ij} &=& -\frac{v^2}{M_N} Y_{(N)}^i Y_{(N)}^j
\end{eqnarray}
since the $N$ ``pairs off'' with only one combination of the neutrinos through the Dirac-like couplings. Here the indices $i,j = 1,2,3$ are the flavour indices corresponding to the three generations of lepton doublets $L^k$. Choosing $M_N\, \sim\, 10^{12}\,\text{GeV}$ (at the seesaw scale),  $\mathbb{Y}_{(N)}\, \sim\, 10^{-1}$, one reproduces the required value of the atmospheric scale.

\subsection{Radiative Contributions}
\label{sec:radcor}

There are quantum corrections to Eq.~(\ref{tree-part}) arising from loop diagrams. Most will simply correspond to renormalization of our (3,1) type-I seesaw mechanism, reproducing the same projective structure that leads to the masslessness of the two lighter neutrinos. These corrections are indicated by a blob in (\ref{eq:nnmassRAW}) given below,
\begin{eqnarray}
\mathcal{M}_{\nu~\rm TOTAL}^{ij}  &=& 
\left[ \begin{array}{ccc}
             & \quad & \\
             & \quad & \\
           \quad
           \parbox{20mm}{
           \begin{fmffile}{radneutmassv2}
           \begin{fmfgraph*}(50,50) 
           \fmfleft{i2,i1} \fmfright{o2,o1}
           \fmf{dashes}{i1,v1}
           \fmfdot{i1}
           \fmflabel{$\left< H \right>$}{i1}
           \fmf{dashes}{v1,o1}
           \fmfdot{o1}
           \fmflabel{$\left< H \right>$}{o1}
           \fmf{dashes,left=0.4,tension=0.4}{v2,v1}
           \fmf{dashes,left=0.4,tension=0.4}{v1,v4}
           \fmf{plain_arrow,label=$\nu$,l.s=left}{i2,v2}
           \fmf{plain_arrow,tension=0.4}{v2,v3}
           \fmf{plain_arrow,tension=0.4}{v4,v3}
           \fmf{plain_arrow,label=$\nu$,l.s=right}{o2,v4}
           \end{fmfgraph*}
           \end{fmffile}} 
           & \quad  &
           \quad\quad
           \parbox{20mm}{ 
           \begin{fmffile}{nnmat24} 
           \begin{fmfgraph*}(50,50)  
           \fmfleft{i1,i2} \fmfright{o1} 
           \fmf{plain_arrow}{v1,i1} \fmflabel{$N$}{i1} 
           \fmf{dashes}{i2,v1} \fmflabel{$\left< H \right>$}{i2} 
           \fmfblob{.2w}{v1}
           \fmf{plain_arrow}{o1,v1}  \fmflabel{$\nu$}{o1} 
           \fmfdot{i2}
           \end{fmfgraph*} 
           \end{fmffile}} \bigskip\quad\quad\quad
            \\
            & \quad &  \\
            & \quad &  \\
            \bigskip\quad\quad\quad\quad
           \parbox{20mm}{ 
           \begin{fmffile}{nnmat20} 
           \begin{fmfgraph*}(50,50)  
           \fmfleft{i1,i2} \fmfright{o1} 
           \fmf{plain_arrow}{i1,v1} \fmflabel{$\nu$}{i1} 
           \fmf{dashes}{i2,v1} \fmflabel{$\left< H \right>$}{i2} 
           \fmfblob{.2w}{v1}
           \fmf{plain_arrow}{v1,o1}  \fmflabel{$N$}{o1} 
           \fmfdot{i2}
           \end{fmfgraph*} 
           \end{fmffile}} 
           \bigskip\quad\quad
             & \quad  &
             \quad\quad
           \parbox{20mm}{ 
           \begin{fmffile}{nnmat26} 
           \begin{fmfgraph*}(50,50)  
           \fmfleft{i1} \fmfright{o1} 
           \fmf{plain_arrow}{i1,v1} \fmflabel{$N$}{i1} 
           \fmfblob{.2w}{v1}
           \fmf{plain_arrow}{o1,v1} \fmflabel{$N$}{o1} 
           \end{fmfgraph*} 
           \end{fmffile}} \quad
        \end{array} \right] \label{eq:nnmassRAW}
\end{eqnarray}

Such corrections only redefine the parameters and do not add a new independent neutrino mass scale, needed in order to account for solar neutrino oscillations. Hence the mechanism generating the latter must be intrinsically different from the N-mediated seesaw. In order to implement this we clone our (3,1) seesaw  mechanism with an independent low-scale mechanism of neutrino mass generation~\cite{Boucenna:2014zba,Cai:2017jrq} of the scotogenic type~\cite{Ma:2006km, Ma:2008cu, Hirsch:2013ola, Hernandez:2013dta, Aranda:2015xoa, Ding:2016wbd, Borah:2016zbd, Wang:2017mcy}. 
This last step is crucial in order to complete the theory with an experimentally testable dark matter sector. \\ \\

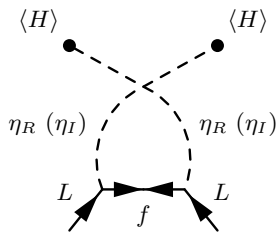
\begin{figure}[!ht]
\parbox{50mm}{
\begin{fmffile}{radneutmass}
\begin{fmfgraph*}(70,70) 
\fmfleft{i2,i1} \fmfright{o2,o1}
\fmf{dashes}{i1,v1}
\fmfdot{i1}
\fmflabel{$\left< H \right>$}{i1}
\fmf{dashes}{v1,o1}
\fmfdot{o1}
\fmflabel{$\left< H \right>$}{o1}
\fmf{dashes,label=$\eta_{R}$ ($\eta_{I}$),left=0.4,tension=0.4}{v2,v1}
\fmf{dashes,label=$\eta_{R}$ ($\eta_{I}$),left=0.4,tension=0.4}{v1,v4}
\fmf{plain_arrow,label=$L$,l.s=left}{i2,v2}
\fmf{plain_arrow,tension=0.4}{v2,v3}
\fmflabel{$f$}{v3}
\fmf{plain_arrow,tension=0.4}{v4,v3}
\fmf{plain_arrow,label=$L$,l.s=right}{o2,v4}
\end{fmfgraph*}
\end{fmffile}}
\bigskip
\caption{Radiative neutrino mass involving dark sector exchange, leading to a calculable solar neutrino mass scale.}
\label{fig:radneutmass}
\end{figure}

In this case there are genuinely calculable quantum corrections arising from Fig.~(\ref{fig:radneutmass}), and involving the exchange of the scalar and fermionic dark messengers $\eta$ and $f$. Though also projective, these corrections can not be renormalized away, and break the degeneracy of neutrino masses intrinsic to the “incomplete” nature of the (3,1) type-I template seesaw mechanism. The total neutrino mass has the structure
\begin{eqnarray}
\mathcal{M}_{\nu~\rm TOT}^{ij} &=& -\frac{v^2}{M_N} Y_{(N)}^i Y_{(N)}^j \,+\, \mathcal{F} \left(m_{\eta^R},m_{\eta^I},M_f \right) M_{f} Y_{(f)}^i Y_{(f)}^j \label{eq:loop-part}
\label{eq:numass}
\end{eqnarray}
where the function $\mathcal{F}$ is a loop function, expressed as the difference of two $B_0$-Veltman functions, namely,
\begin{eqnarray}
\mathcal{F} \left(m_{\eta^R},m_{\eta^I},M_f \right) &=& \frac{1}{32\pi^2}
 \left[ \frac{m_{\eta^R}^2 \log \left({M_{f}^2/m_{\eta^R}^2}\right)}{M_{f}^2-m_{\eta^R}^2} -
        \frac{m_{\eta^I}^2 \log \left({M_{f}^2/m_{\eta^I}^2}\right)}{M_{f}^2-m_{\eta^I}^2}
\right] \label{eq:Cfactor}
\end{eqnarray}
One sees that these are ``genuine'' corrections with a different structure, that can lift the degeneracy and thus generate the solar neutrino mass splitting. Moreover, these are parametrically unrelated to the tree level masses. As shown above in (\ref{eq:nnmassRAW}), it is simple to understand in diagrammatic form the different structure of the tree and loop contributions. Moreover, given the projective nature of both terms in Eq.~(\ref{eq:numass}) one of the three neutrinos remains massless.

We now give a simple estimate of the relevant parameters required in order to account for the observed neutrino oscillations. The solar and the atmospheric square mass differences can be estimated by each of the eigenvalues of the tree level and radiative contributions to the neutrino mass matrix in Eq.~(\ref{eq:numass})
  \begin{eqnarray}
\label{eq:scales}
  \Delta m_{\rm ATM}^2 &\sim& \left( \frac{v^2}{M_N} \mathbb{Y}_{(N)}^2 \right)^2 \, ,\quad \Delta m_{\rm SOL}^2\,\sim\, \left(\frac{1}{32\pi^2}\right)^2\left( \frac{\lambda_5 v^2}{M_f^2 - m_{\eta}^{(R)2}} M_f \mathbb{Y}_{(f)}^2 \right)^2.
  \end{eqnarray}
The expression for $\Delta m_{\rm SOL}^2$ is obtained through Eqs.~(\ref{eq:etaR}) and (\ref{eq:etaI}), where we take $M_f^2,\, m_{\eta}^{(R)2},\, M_f^2 - m_{\eta}^{(R)2} \gg \lambda_5 v^2$, defining $\mathbb{Y}_{(a)}^2 = \left( Y_{(a)}^{e} \right)^2\,+\, \left( Y_{(a)}^{\mu} \right)^2\,+\, \left( Y_{(a)}^{\tau} \right)^2$ for $a=f,N$, which summarizes the role of the Yukawa couplings. \\[-.2cm]
 
Hence, the ratio between the solar and atmospheric square mass differences can be estimated by:
 \begin{eqnarray}
  \frac{\Delta m_{\rm SOL}^2}{\Delta m_{\rm ATM}^2} &\sim& \left( \frac{1}{32\pi^2} \right)^2 \lambda_5^2 \left( \frac{M_N M_f}{M_f^2 - m_{\eta}^{(R)2}} \right)^2 \left( \frac{\mathbb{Y}_{(f)}^{2}}{\mathbb{Y}_{(N)}^{2}} \right)^2
  \label{eq:ratio}
  \end{eqnarray}

A priori, the mass of the right handed neutrino $M_N$ and that of the dark sector fermion $M_f$ are two independent parameters that, in principle, can have very different values.
As an example, the case of scalar WIMP dark matter can be realized by choosing $M_N\, \sim\, 10^{12}\,\text{GeV}$, $M_f\sim 10^{4}\,\text{GeV}$, $m_{\eta}^{(R)}\, \sim\, 10^{3}\,\text{GeV}$, $\mathbb{Y}_{(N)}\, \sim\, 10^{-1}$, $\mathbb{Y}_{(f)} \sim 10^{-4}$, 
it can easily fit the solar and atmospheric scales as well as their ratio in Eq.~(\ref{eq:sol-atm}), as long as one takes an adequately small value for $\lambda_5$.
The latter indicates symmetry protection since, as $\lambda_5 \to 0$, lepton number is recovered in the radiative sector.
However, the masses of $N$ and $f$ can be quite similar to each other. For example, keeping other parameters as before one can also have $M_N \approx M_f \approx 10^4$ GeV for $\mathbb{Y}_{(N)}\, \sim\, 10^{-5}$ and $\mathbb{Y}_{(f)} \sim 10^{-4}$. An alternative benchmark with intermediate mass values $M_N \approx M_f \approx 10^6$ GeV is obtained for $\mathbb{Y}_{(N)}\, \sim\, 10^{-4}$ and $\mathbb{Y}_{(f)} \sim 10^{-3}$. Clearly, many other choices for masses and couplings can be found using Eq.~\eqref{eq:ratio}. The upshot of the discussion is that one can easily fit Eq.~(\ref{eq:sol-atm}), by making reasonable parameter choices.


\section{Lepton Flavour Violation}

Before closing, let us comment briefly on lepton flavour violation. 
Our model contains sources of lepton flavour violation, arising 
from the new Yukawa couplings $Y_f$ and $Y_N$ associated to the ``scotogenic'' 
and ``seesaw'' sectors of the theory.
These lead to processes such as $\mu \to e\gamma$, $\mu \to 3e$ and 
muon-to-electron conversion in nuclei.
For adequate $M_N$ and $Y_N$ values able to reproduce the required value of 
the neutrino mass scales in Eq.~\ref{eq:scales}, the ``scotogenic'' 
contribution is by far the dominant one, hence we neglect the others
\footnote{ For instance, the pure seesaw contribution to $\mu \to e\gamma$ ranges between $\mathcal{O} (10^{-50}) - \mathcal{O} (10^{-34})$ for $M_N \sim 10^{12} - 10^{4}$ GeV, respectively.}.
%
%
The lepton flavour violating branching fractions for $\mu \to e\gamma$ and $\mu \to 3e$ are 
given as
\begin{eqnarray}
  BR\left( \mu \to  e\gamma \right) &\approx& \frac{\alpha m_\mu}{\left(8\pi\right)^4\Gamma_t} \left( \frac{m_\mu}{m_\eta} \right)^4 \left( Y_f^1 Y_f^2\right)^2\, \left( F \left( \frac{M_f^2}{m_\eta^2} \right) \right)^2 \\
  BR\left( \mu \to  3e \right) &\approx& \frac{\alpha^2 m_\mu}{\left(8\pi\right)^5\Gamma_t} \left( \frac{m_\mu}{m_\eta} \right)^4 \left( Y_f^1 Y_f^2 \right)^2\, G \left( \frac{m_\mu}{m_e} \right) \left( F \left( \frac{M_f^2}{m_\eta^2} \right) \right)^2 
 \end{eqnarray}

In these expressions $\Gamma_t$ is the total muon decay width, which is 
approximately $3\cdot 10^{-19}\, \text{GeV}$, while the functions $F$ and $G$ have the form
\begin{eqnarray}
  F(x) &=& \frac{2\,-\,9x\,+\,18x^2\,-\,11x^3\,+\,6x^3\log\left(x\right)}{6\left(1\,-\,x\right)^4} \\
  G(x) &=& \frac{16}{3}\log\left( x\right)\,-\,\frac{22}{3}
 \end{eqnarray}

Notice the presence of the $\alpha$ factor in addition to the phase space suppression in the $\mu \to  3e $ decay.
Owing to this suppression the most stringent constraints on our model will come  from the $\mu \to  e \gamma $ decay.

\begin{figure}[htbp!]
\includegraphics[width=.7\textwidth]{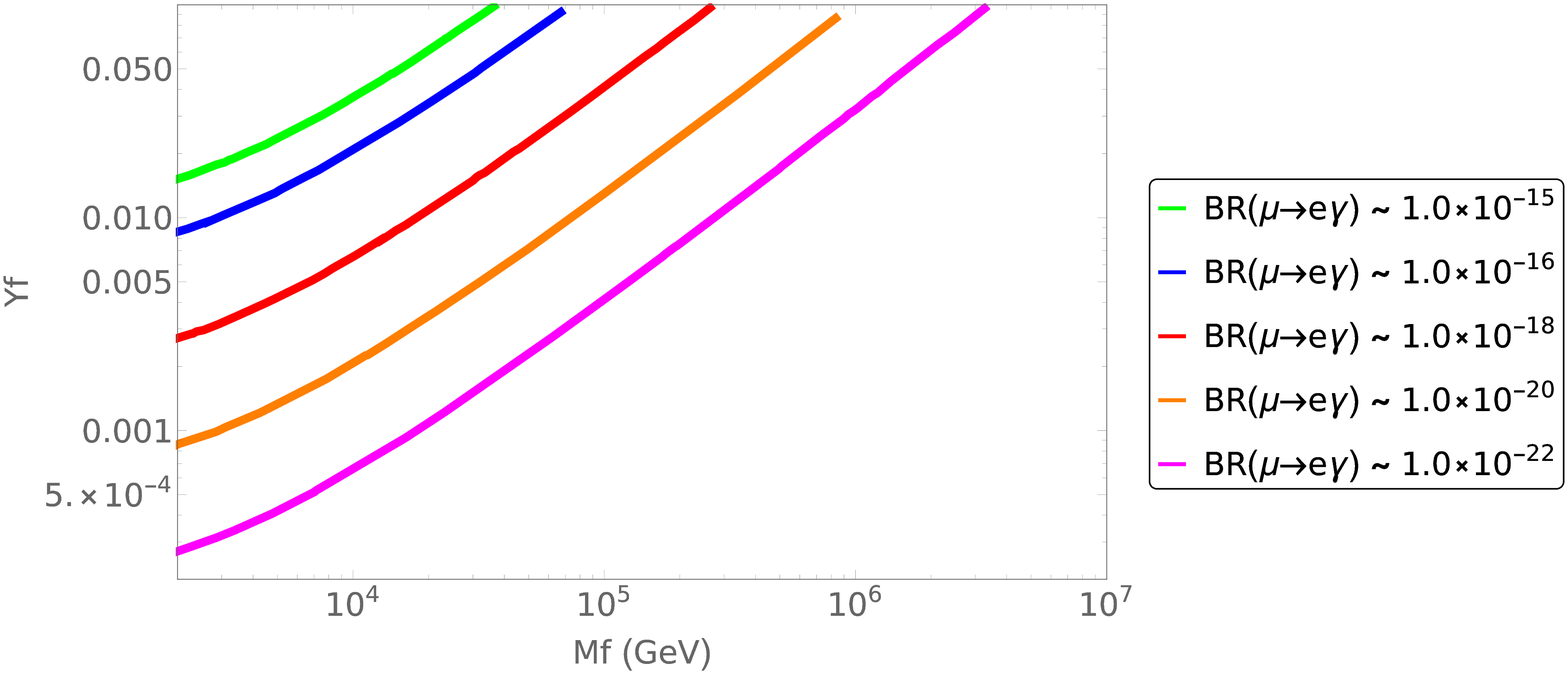}
\caption{\footnotesize Isocontour plots for $BR\left( \mu \to  e\gamma \right)$ in the $M_f$ vs $Y_f$ plane.}
\label{fig:CPBRmueg}
\end{figure}

For illustration, we show in Fig.~\ref{fig:CPBRmueg} the results for the  $\mu \to  e\gamma$ branching ratio in our model. The contours have a fixed value of the branching fraction, as indicated on the right.

One can see that the expected values for the lepton flavour violating branching fractions for the benchmarks used in sec.~\ref{sec:radcor} fall well within the experimental upper bounds for 
$BR\left( \mu \to  e\gamma \right)\sim 4.2\cdot 10^{-13}$ and $BR\left( \mu \to  3e \right)\sim 1.0\cdot 10^{-12}$~\cite{Tanabashi:2018oca}.
A dedicated study of \lfv within our model is outside the scope of this paper, and related material for the scotogenic model can be found in Refs.~\cite{Toma:2013zsa,Vicente:2014wga}. 


\section{Discussion and outlook}

Having proposed the minimal setup for neutrino mass and dark matter, some comments are in order. 
First we note that neutrino masses arise from the interplay of tree level and quantum corrections, in a way reminiscent of supersymmetric models with bilinear breaking of R-parity. However, our theory is much simpler and, by construction, more complete, as it naturally incorporates the existence of a dark matter particle.
The setup is minimalistic, though realistic, having effectively two messengers, the tree-level seesaw atmospheric messenger, and the radiative dark messenger responsible for inducing solar neutrino oscillations. Given the fact that one of the three neutrinos is massless in our model, and given the current information on the neutrino oscillation parameters~\cite{deSalas:2017kay}, one finds that the neutrinoless double beta decay (\znbb) amplitude never vanishes~\cite{Reig:2018ztc}. Indeed, the ``incomplete messenger'' structure of our model translates into a lower bound for \znbb. This holds even if the ordering of light neutrino mass eigenvalues is of the normal type, as currently preferred by the oscillation data.

Concerning dark matter we stress that the model has a built-in dark matter candidate that can be detected by nuclear recoil experiments such as Xenon-1T~\cite{Aprile:2018dbl,Aprile:2017iyp}. Alternative options as well as extended scenarios can, however, easily be envisaged. In fact, possible variations of the original scotogenic model suggest new realizations of the idea proposed here.
New charged fermions odd under the $\mathbb{Z}_2$ symmetry can be added instead of, or in addition to, the neutral fermion $f$ in table~\ref{tab:Fields}. For example, one can replace the fermion $f$ by a hyperchargeless isotriplet $\Sigma$, as in~\cite{Ma:2008cu} or add this triplet field altogether, as in~\cite{Hirsch:2013ola}. The dark matter candidate could then be an isotriplet-isosinglet fermion mixture or a scalar boson, generating new radiative contributions to the effective neutrino mass matrix in Eq.~(\ref{eq:numass}). There will be a wider parameter space for obtaining the correct relic density~\cite{Hirsch:2013ola}, as well as a consistent high energy behavior of the required $\mathbb{Z}_2$ symmetry~\cite{Merle:2016scw}. 
Likewise, a rich pattern of lepton flavour violating processes, such as $\mu \to e+\gamma$, $\mu \to 3e$ and $\mu-e$ conversion in nuclei is expected, 
as well as new possibilities concerning the issue of leptogenesis. In short, despite its simplicity and minimality, the model exhibits a rich phenomenological potential which deserves dedicated study. We intend to explore them in a future work.

\section{Summary}
\label{sec:summary-discussion}

We have proposed the simplest scotogenic extension of the minimum type-I seesaw scenario as a way to explain the hierarchy observed between the solar and atmospheric neutrino mass scales, Eq.~(\ref{eq:sol-atm}), as well as to harbor a viable dark matter candidate. The template is the (3,1) type-I seesaw mechanism proposed in~\cite{Schechter:1980gr} characterized by an ``incomplete'' lepton sector consisting of a single ``right-handed'' neutrino. The setup is the minimal way to generate neutrino masses consistently, as well as the WIMP dark matter candidate. It gives a clear interpretation of the solar and atmospheric oscillation scales and the associated messengers. In the tree-level approximation one generates only the atmospheric scale, from the exchange of the single right-handed neutrino present in the (3,1) type-I seesaw mechanism, while the solar scale arises from calculable loop corrections mediated by the dark sector. 

\begin{acknowledgments}

Work supported by the Spanish grants FPA2017-85216-P and SEV-2014-0398, and PROMETEOII/2018/165 (Generalitat  Valenciana). NR was funded by proyecto FONDECYT Postdoctorado Nacional (2017) num. 3170135. He would like to thank all the people at AHEP group at IFIC for the hospitality. We thank Martin Hirsch, Christoph Termes and Avelino Vicente for useful input.

\end{acknowledgments}

\bibliographystyle{utphys}
\bibliography{bibliography}

\end{document}